\documentclass[sor,jor,graphicx,superscriptaddress, preprint,showkeys]{revtex4-1}

\usepackage{graphicx}
\usepackage{amssymb}

\usepackage{lineno}

\usepackage{hyperref}
\usepackage{subfig} 
\usepackage{siunitx} 
\usepackage{booktabs} 
\usepackage{amsmath} 
\usepackage{graphicx, amssymb, wasysym,epsf, graphicx, enumitem, tikz}
\usepackage{chemformula}
\usepackage{textcomp} 

\begin{document}


\title{Viscosity of protein-stabilised emulsions: contributions of components and development of a semi-predictive model} 



\author{Marion Roullet}
\affiliation{Unilever R\&D Colworth, Sharnbrook, Bedford, MK44 1LQ, UK}
\affiliation{School of Physics and Astronomy, University of Edinburgh, Peter Guthrie Tait Road, Edinburgh, EH9 3FD, UK}

\author{Paul S. Clegg}
\affiliation{School of Physics and Astronomy, University of Edinburgh, Peter Guthrie Tait Road, Edinburgh, EH9 3FD, UK}

\author{William J. Frith}
\affiliation{Unilever R\&D Colworth, Sharnbrook, Bedford, MK44 1LQ, UK}


\date{\today}

\begin{abstract}
Protein-stabilised emulsions can be seen as mixtures of unadsorbed proteins and of protein-stabilised droplets. To identify the contributions of these two components to the overall viscosity of sodium caseinate o/w emulsions, the rheological behaviour of pure suspensions of proteins and droplets were characterised, and their properties used to model the behaviour of their mixtures. These materials are conveniently studied in the framework developed for soft colloids. Here, the use of viscosity models for the two types of pure suspensions facilitates the development of a semi-empirical model that relates the viscosity of protein-stabilised emulsions to their composition.
\end{abstract}

\pacs{}
\keywords{Viscosity; Emulsions; Sodium caseinate; Soft colloids; Mixtures}

\maketitle 


\section{Introduction}

Despite their complexity, food products can be conveniently studied from the perspective of colloid science \cite{mezzenga:2005}. In the last three decades, research in the field of food colloids has led to major advances  in understanding their structure over a wide range of lengthscales \cite{dickinson:2011}, which has proved key to developing a good control of their flavour and texture properties \cite{vilgis:2015}.

\begin{figure}[htb]
	\begin{center}
		\includegraphics[width=0.8\textwidth]{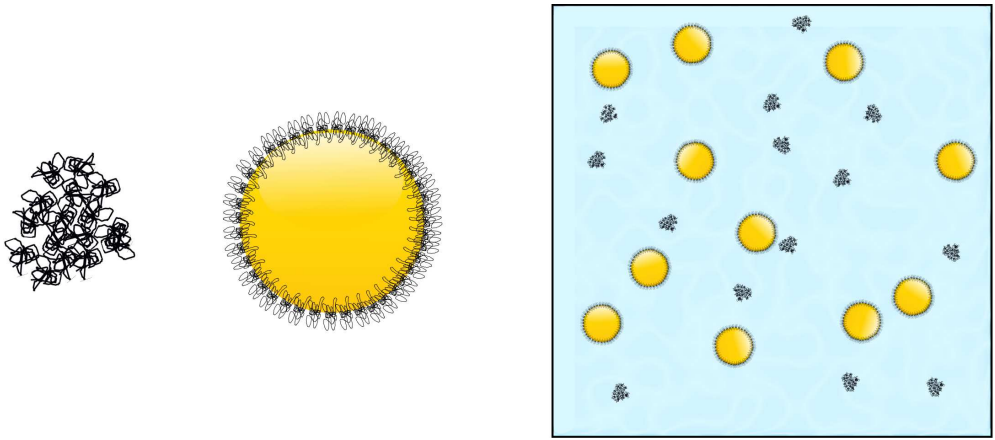}
	\end{center}
	\caption{Illustration of a protein assembly, protein-stabilised droplet, and protein-stabilised emulsion seen as a mixture of droplets and un-adsorbed proteins.
	}
	\label{Fig:CartoonProtDrop}
\end{figure}

Many food products such as mayonnaise, ice cream, and yogurt involve protein-stabilised emulsions either during their fabrication or as the final product. Proteins have particularly favourable properties as emulsifiers because of their ability to strongly adsorb at oil/water interfaces and to stabilise oil droplets by steric and electrostatic repulsion. However, proteins do not completely adsorb at the interface, leaving a residual fraction of protein suspended in the continuous phase after emulsification \cite{srinivasan:1996,srinivasan:1999}. Protein-stabilised emulsions are thus mixtures of protein-stabilised droplets and suspended proteins, as illustrated in Figure~\ref{Fig:CartoonProtDrop}. Understanding the contributions of these two components to the properties of the final emulsion remains a challenge. 

When considered separately, the droplets in protein-stabilised emulsions can be considered as colloidal particles with some degree of softness. It is thus possible to compare the rheological properties of protein-stabilised emulsions to other types of soft particle suspension and to model their behaviour. From a theoretical point of view, particles, colloidal or not, can be described as soft if they have the ability to change size and shape at high concentration \cite{vlassopoulos:2014}. Such a definition covers a striking variety of systems, including gel microparticles \cite{adams:2004,shewan:2015}, microgels \cite{cloitre:2003,tan:2005}, star polymers \cite{roovers:1994, winkler:2014} and block co-polymer micelles \cite{lyklema:2005:4}. These systems have been the focus of many studies in the last two decades, however one major challenge to comparing the behaviour  of such diverse materials is the availability of a well defined volume fraction $\phi$ for the suspensions. 

To overcome the challenge of defining the volume fraction of soft colloids, a common approach is to use an effective volume fraction $\phi_{eff}$ proportional to the concentration $c$, $\phi_{eff}=k_0\times c$, where $k_0$ is a constant indicating the voluminosity of the soft particle of interest, usually determined in the dilute or semi-dilute regime. Such a definition for $\phi_{eff}$ does not take into account the deformation or shrinking of the particle at high concentrations, high values of  ($\phi_{eff}>1$) can thus be reached. $k_0$ can be estimated using osmometry \cite{farrer:1999}, light scattering \cite{vlassopoulos:2001} or viscosimetry \cite{tan:2005,roovers:1994,boulet:1998}. In this study, $k_0$ was estimated, for each individual component of the emulsions, by modelling the relative zero-shear viscosity $\eta_0/\eta_s$ behaviour of the pure suspensions in the semi-dilute regime with Batchelor equation for hard spheres \cite{batchelor:1977}: 
\begin{equation}
\label{Eq:Batchelor}
\frac{\eta_0}{\eta_s} = 1+2.5 \phi_{eff}+6.2 \phi_{eff}^2
\end{equation}

Sodium caseinate is used here to stabilise emulsions as a case-study, because of its outstanding properties as a surface-active agent and stabiliser, and because it is widely used in industry. Sodium caseinate is produced by replacing the calcium in native milk casein micelles, with sodium, to increase its solubility \cite{dalgleish:1988}, a process which also leads to the disruption of the micelles. It has been established that sodium caseinate is not present as a monomer in suspension, but rather in the form of small aggregates \cite{lucey:2000}. The exact nature of the interactions in play in the formation of these aggregates is not well-known but they have been characterised as elongated and their size estimated to be around $\SI{20}{\nano\metre}$ \cite{farrer:1999,lucey:2000,huppertz:2017}. Some larger aggregates can also form in presence of residual traces of calcium or oil from the original milk, however these only represent a small fraction of the protein \cite{dalgleish:1988}. The viscosity behaviour of sodium caseinate as a function of concentration shows similarities with hard-sphere suspensions at relatively small concentrations, but at higher concentrations, over $c>\SI{130}{\gram\per\liter}$, the viscosity continues to increase with a power-law rather than diverging \cite{farrer:1999,pitowski:2008} as would be expected for a hard sphere suspension \cite{faroughi:2014}.

In this study, the rheology of protein-stabilised emulsions is examined within the framework of soft colloidal particles. Modeling proteins in this way ignores protein-specific elements, such as surface hydration, conformation changes, association, and surface charge distribution \cite{sarangapani:2013,sarangapani:2015}, but it provides a convenient theoretical framework to separate and discuss the contributions of both sodium caseinate and the droplets to the viscosity of emulsions. Similarly, protein-stabilised droplets can be seen as comprising an oil core and a soft protein shell \cite{bressy:2003}, allowing for a unifying approach for both components of the emulsions.

The aim of this study is to present a predictive model of the viscosity of protein-stabilised emulsions, that takes into account the presence and behaviour of both the protein stabilised droplets and the unadsorbed protein. A first step is to characterise separately the flow behaviour and viscosity of suspensions of purified protein-stabilised droplets, and of protein suspensions over a wide range of concentrations. This also allows a critical assessment of the soft colloidal approach. These components are then combined to form mixtures of well-characterised composition and their viscosity is compared to a semi-empirical model. Because they are well dispersed, most of the suspensions and emulsions display a Newtonian behavior at low shear, with shear thinning at higher shear-rates. In this context, we model the concentration dependence of zero-shear viscosity and the shear thinning behaviour separately to confirm the apparent colloidal nature of the components of the emulsions and protein suspensions.

\section{Materials \& Methods}

\subsection{Preparation of protein suspensions}
Because of its excellent ability to stabilise emulsions, sodium caseinate (Excellion S grade, spray-dried, kindly provided by DMV, Friesland Campina, Netherlands), was used in this study.  It was further purified by first suspending it in deionised water, at $5-9 \%$ (w/w), and then by mixing thoroughly with a magnetic stirrer for $\SI{16}{\hour}$. After complete dispersion, a turbid suspension was obtained, which was centrifuged at $\num{40000}\times$g (Evolution RC, Sorvall with rotor SA 600, Sorvall and clear $\SI{50}{\milli\liter}$ tubes, Beckmann) for $\SI{4}{\hour}$ at $\SI{21}{\degreeCelsius}$. Subsequently, the supernatant, made of residual fat contamination, and the sediment were separated from the suspension, that was now clearer. The solution was then filtered using a $\SI{50}{\milli\liter}$ stirred ultra-filtration cell (Micon, Millipore) with a $\SI{0.45}{\micro\meter}$ membrane (Sartolon Polyamid, Sartorius). In order to avoid spoilage of the protein solution 0.05\% of ProClin 50 (Sigma Aldrich) was added. The suspension at $5\%$ (w/w) was then diluted to the required concentration. Concentrated suspensions of sodium caseinate were prepared by evaporating a stock solution of sodium caseinate at $\SI{5}{\percent}$(w/w), prepared following the previous protocol, using a rotary evaporator (Rotavapor R-210, Buchi). Mild conditions were used to avoid changing the structure of the proteins: the water bath was set at $\SI{40}{\degreeCelsius}$ and a vacuum of $\SI{45}{\milli\bar}$ was used to evaporate water. The concentration of all the suspensions after purification was estimated by refractometry, using a refractometer RM 50 (Mettler Toledo), LED at $\SI{589.3}{\nano\metre}$ and a refractive index increment of $dn/dc =\SI[separate-uncertainty=true]{0.1888(00033)}{\milli\liter\per\gram}$ \cite{zhao:2011}.

\begin{figure}[htb]
	\begin{center}
		\includegraphics[width=0.9\textwidth]{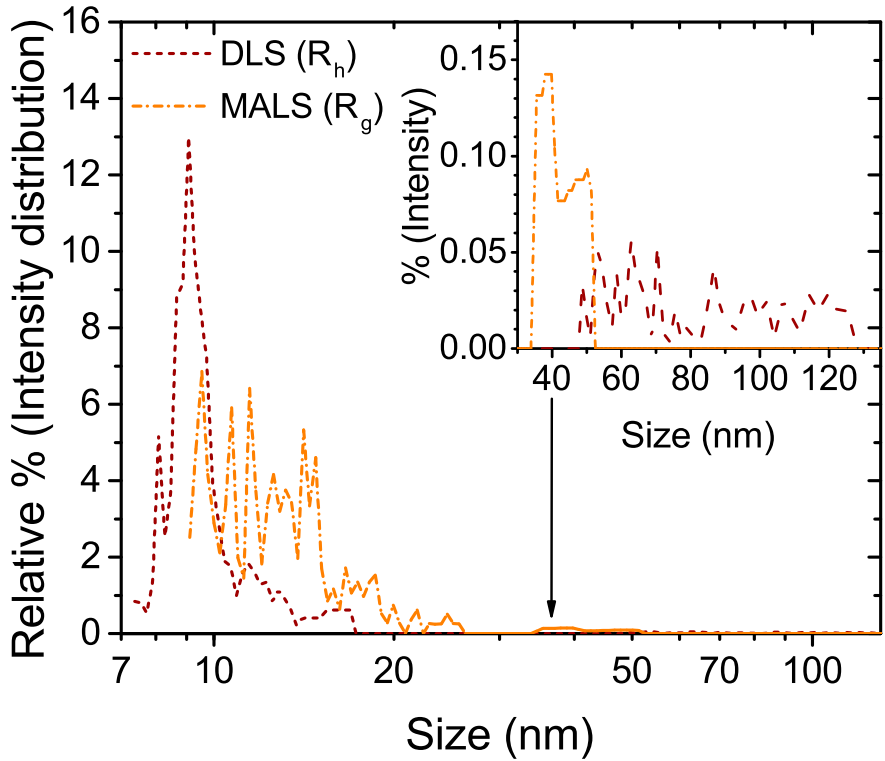}
	\end{center}
	\caption{Size distributions of sodium caseinate after the purification protocol. The sample was fractionated by Asymetric Flow Field Flow Fractionation (kindly performed by PostNova Analytics Ltd), and the sizes were measured online by Dynamic Light Scattering (dot line, red) and by Multi Angle Light Scattering (dash dot line, orange). The relative percentage of each class is weighted by the intensity of the scattered light. The inset is a zoom of the small fraction of proteins that are present as larger aggregates 
	}
	\label{Fig:SizeProtFFF}
\end{figure}

Size analysis by Flow Field Fractionation (kindly performed by PostNova Analytics Ltd) showed that the resulting suspensions of sodium caseinate were composed of small aggregates of a hydrodynamic radius of $\SI{11}{\nano\metre}$ at $\SI{96}{\percent}$, while the remaining $\SI{4}{\percent}$ formed larger aggregates with a wide range of sizes (hydrodynamic radii from $\SI{40}{\nano\metre}$ to $\SI{120}{\nano\metre}$) as shown in Figure~\ref{Fig:SizeProtFFF}.

\subsection{Preparation of emulsions}
Nano-sized caseinate-stabilised droplets were prepared in two steps. First, the pre-emulsion was produced by mixing $\SI{45}{\milli\gram\per\milli\liter}$ sodium caseinate solution (prepared as detailed previously) with glyceryl trioctanoate ($\rho=\SI{0.956}{\gram\per\milli\liter}$, Sigma Aldrich) at a weight ratio 4:1 using a rotor stator system (L4R, Silverson). This pre-emulsion was then stored at $\SI{4}{\degreeCelsius}$ for $\SI{4}{\hour}$ to reduce the amount of foam. It was then passed through a high-pressure homogeniser (Microfluidizer, Microfluidics) with an input pressure of $\SI{5}{\bar}$, equivalent to a pressure of $\approx\SI{1000}{\bar}$ in the micro-chamber, three times consecutively. After 3 passes, a stationary regime was reached where the size of droplets could not be reduced any further. This protocol for emulsification produced droplets of radius around $\SI{110}{\nano\metre}$ as measured by Dynamic Light Scattering (Zetasizer, Malvern) and $\SI{65}{\nano\metre}$ by Static Light Scattering (Mastersizer, Malvern). 

Because not all the protein content was adsorbed at the interface, an additional centrifugation step was required to separate the droplets from the continuous phase of protein suspension. This separation was performed by spinning the emulsion at $\num{235000}\times$g with an ultra-centrifuge (Discovery SE, Sorvall, with fixed-angle rotor $45$Ti, Beckmann Coulter) for $\SI{16}{\hour}$ at $\SI{21}{\degreeCelsius}$. The concentrated droplets then formed a solid layer at the top of the subnatant that could be carefully removed with a spatula. The subnatant containing proteins and some residual droplets was discarded. The drying of a small fraction of the concentrated droplet layer and the weighing of its dry content yielded a concentration of the droplet paste of $\SI[separate-uncertainty=true]{0.519(0008)}{\gram\per\milli\liter}$, so the concentration in droplets of all the suspensions were derived from the dilution parameters. Only one centrifugation step was employed to separate the droplets from the proteins, as it was felt that further steps may lead to protein desorption and coalescence.
The pure nano-sized droplets were then re-dispersed at the required concentration, in the range $\num{0.008}$ to $\SI{0.39}{\gram\per\milli\liter}$ in deionised water for $1$ to $\SI{30}{\minute}$ with a magnetic stirrer.

\subsection{Preparation of mixtures}
To prepare emulsions with a controlled concentration of proteins in suspension, the concentrated droplets were re-suspended in a protein suspension at the desired concentration using a magnetic stirrer and a stirring plate for $\SI{5}{\minute}$ to $\SI{2}{\hour}$.

\subsection{Viscosity measurements}
Rotational rheology measurements were performed using a stress-controlled MCR 502 rheometer (Anton Paar) and a Couette geometry (smooth bob and smooth cup, $\SI{17}{\milli\liter}$ radius) at $\SI{25}{\degreeCelsius}$. For each sample, three measurements are performed and averaged to obtain the flow curve. The values of viscosity on the plateau at low shear are averaged to determine the zero-shear viscosity. Viscosity measurements were performed at different concentrations for protein suspensions, protein-stabilised droplet suspensions, and mixtures.

\section{Results \& Discussion}
\label{S:1}
In order to study the rheological behaviour of protein-stabilised emulsions, the approach used here is to separate the original emulsion into its two components, namely un-adsorbed protein assemblies and protein-coated droplets, and to characterise the suspensions of each of these components. Despite their intrinsic complexity due to their biological natures, random coil proteins such as sodium caseinate can conveniently be considered as colloidal suspensions, as we demonstrate in the discussion below.

\subsection{Viscosity of suspensions in the semi-dilute regime: determination of volume fraction}
The weight concentration (in $\SI{}{\gram\per\milli\liter}$) is a sufficient parameter to describe the composition in the case of one suspension, but only the use of the volume fraction of the suspended particles allows meaningful comparisons between protein assemblies and droplets. In the framework of soft colloids, the effective volume fraction $\phi_{eff}$ of a colloidal suspension can be determined by modelling the viscosity in the semi-dilute regime with a hard-sphere model.

\begin{figure}[htb]
	\centering
		\includegraphics[width=0.9\textwidth]{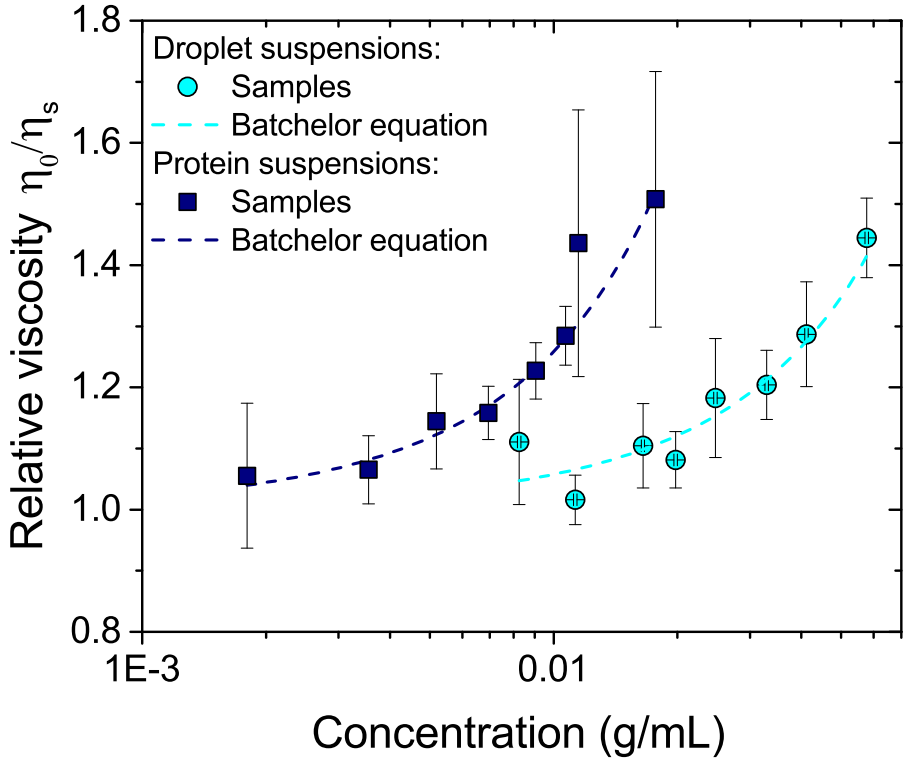}
		
		\caption{Relative viscosity of sodium caseinate suspensions ($\square$, navy blue) and sodium caseinate-stabilised droplets ($\bigcirc$, cyan) as a function of the concentration of dispersed material. The lines denote Batchelor model for hard spheres in the dilute regime, Equation\ref{Eq:Batchelor}.}
	\label{Fig:DiluteRegimeConc}
\end{figure}

The relative zero-shear viscosities of semi-dilute samples are displayed in Figure~\ref{Fig:DiluteRegimeConc} as a function of the mass concentration of protein or droplets (viscosity data at the full range of concentration can be found in Figure S2 in the supplementary material). As can be seen, protein suspensions reach a higher viscosity at a lower weight fraction than droplet suspensions. This is because the protein is highly hydrated and swollen, and so occupies a greater volume per unit mass than do the droplets, where the main contributor to the occupied volume is the oil core. 

The viscosity behaviour of each type of suspension in the semi-dilute regime can be described by a theoretical model such as Batchelor's equation \cite{batchelor:1977}, Equation~\ref{Eq:Batchelor} as a function of the volume fraction $\phi$. This involves assuming that the particles in the suspension of interest do not have specific interparticle interactions or liquid interfaces in this regime, and can be accurately described as hard spheres.

In addition, as a first approximation, the effective volume fraction $\phi_{eff}$ of soft particles in suspension is assumed to be proportional to the weight concentration $c$:
\begin{equation}
\label{Eq:EffPhi_proportional}
\phi_{eff}=k_0 \times c
\end{equation}
where $k_0$ is a constant expressed in $\SI{}{\milli\liter\per\gram}$.
This equation is combined with Equation~\ref{Eq:Batchelor} in order to obtain an expression for the viscosity as a function of the concentration. When applied to experimental viscosity values for suspensions of protein or droplets at concentrations in the semi-dilute regime, such an expression allows estimation of $k_0$. The effective volume fraction $\phi_{eff}$ of the suspensions can then be calculated using Equation~\ref{Eq:EffPhi_proportional}.

When fitted to the viscosity data for pure  sodium caseinate and pure droplets, as described above, Equation~\ref{Eq:Batchelor} gives satisfactory fits as shown in Figure~\ref{Fig:DiluteRegimeConc}. The resulting  values for $k_0$ are, for protein suspensions, $k_{0,prot}=\SI[separate-uncertainty=true]{8.53(23)}{\milli\liter\per\gram}$, and for droplet suspensions, $k_{0,drop}=\SI[separate-uncertainty=true]{2.16(13)}{\milli\liter\per\gram}$. 

 The protein result is in reasonable agreement with previous results, where determinations of the volume fraction using the intrinsic viscosity gave $\phi_{eff,prot} =  \num[separate-uncertainty=true]{6.4}~c$ \cite{pitowski:2008} and  $\phi_{eff,prot} = \num[separate-uncertainty=true]{6.5(5)}~c$ \cite{huppertz:2017}, while osmometry measurements (at a higher temperature) gave $\phi_{eff,prot}=\num{4.47}~c$ \cite{farrer:1999}.
For droplet suspensions, $k_{0,drop}$ corresponds to the voluminosity of the whole droplets. If these were purely made of a hard oil core, their voluminosity would be $1/\rho_{oil}=\SI{1.05}{\milli\liter\per\gram}$. The higher value observed can be attributed to the layer of adsorbed proteins at the surface of the droplets. This is an indication that the nano-sized droplets can be modelled as core-shell particles.

These results make it possible to calculate the effective volume fractions $\phi_{eff}$ of both types of suspensions, which is a necessary step to allowing their comparison. It is however important to keep in mind that $\phi_{eff}$ is an estimate of the volume fraction using the hard sphere-assumption, which is likely to break down as the concentration is increased, where deswelling, deformation and interpenetration of the particles may occur \cite{vlassopoulos:2014}.

\subsection{Modelling the viscosity behaviours of colloidal suspensions}

In order to identify the contributions of the components to the viscosity of the mixture, it is important to characterise the viscosity behaviours of the pure suspensions of caseinate-stabilised nano-sized droplets and of sodium caseinate. This is achieved by modelling the volume fraction dependence of the viscosity with equations for hard and soft colloidal particles.

\subsubsection{Suspensions of protein-stabilised droplets} \label{Sec:ViscoDropModel}

\begin{figure}[htb]
	\begin{center}
		\includegraphics[width=0.8\textwidth]{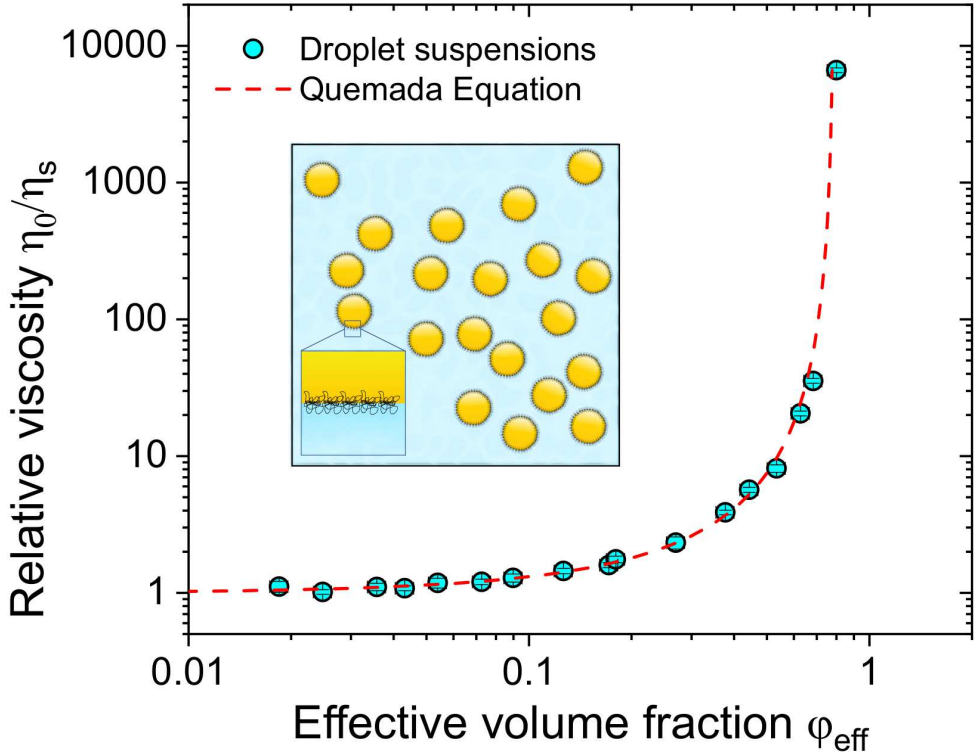}
	\end{center}
	\caption{Relative viscosity of sodium caseinate-stabilised droplets ($\circ$, cyan) as a function of the effective volume fraction. The red dashed line denotes Quemada equation for hard spheres, Equation~\ref{Eq:QuemadaHS} with $\phi_m =\num{0.79}$}
	\label{Fig:ViscoDropQuemada}
\end{figure}

The viscosity of protein-stabilised droplet suspensions is displayed in  Figure~\ref{Fig:ViscoDropQuemada}. A sharp divergence is observed at high volume fraction and this behaviour is typical of hard-sphere suspensions \cite{dekruif:1985}. It is thus appropriate to use one of the relationships derived for such systems to model the viscosity behaviour of droplet suspensions. 

Amongst the multiple models for the viscosity of hard-sphere suspensions that have been proposed over time, the theoretical model developed by Quemada \cite{quemada:1977} is used in this work:
\begin{equation}
\label{Eq:QuemadaHS}
\frac{\eta_0}{\eta_s} = \left(1-\frac{\phi}{\phi_m}\right)^{-2} 
\end{equation}
Where the parameter $\phi_m$ is the maximum volume fraction at which the viscosity of the suspension diverges:
\begin{equation}
\label{Eq:DivergenceVisco}
\lim_{\phi\to\phi_m} \frac{\eta_0}{\eta_s}=\infty
\end{equation}

The Quemada model fits remarkably well to the experimental data of the relative viscosity $\frac{\eta_0}{\eta_s}$ of suspensions of droplets. The value for the maximum volume fraction is found to be $\phi_m  =\num[separate-uncertainty=true]{0.79(2)}$. 
Despite the similarity in viscosity behaviour between the droplet suspensions and hard-sphere suspensions, the maximum volume fraction found here is considerably higher than the theoretical value of $\phi_m=\phi_{rcp}=\num{0.64}$ for randomly close-packed hard spheres.

A possible explanation for this discrepancy is the polydispersity of the droplet suspension. Indeed, random close-packing is highly affected by the size distribution of the particles, as smaller particles can occupy the gaps between larger particles \cite{farris:1968}. In a recent study, Shewan and Stokes modelled the viscosity of hard sphere suspensions using a maximum volume fraction predicted by a numerical model developed by Farr and Groot \cite{shewan:2015b, farr:2009}, which allows the maximum volume fraction of multiple hard-sphere suspensions to be predicted from their size distribution.

Here, the same approach is used with the size distributions of the protein-stabilised droplets obtained from both the Mastersizer and the Zetasizer. The numerically estimated random close-packing volume fraction $\phi_{rcp}$ is close for both size distributions, and its value is $\phi_{rcp}=\num{0.68}$. Although this is a higher maximum volume fraction than for a monodisperse hard-sphere suspension, it is still considerably lower than the experimental value, $\phi_m=\num{0.79}$. Such a high random close-packing fraction can be achieved numerically only if a fraction of much smaller droplets is added to the distribution obtained by light scattering. The hypothesis of the presence of small droplets, undetectable by light scattering without previous fractionation is supported by the observation of such droplets upon fractionation of a very similar emulsion in a previous study \cite{dalgleish:1997}.

It is also possible that other mechanisms than the polydispersity come into play at high volume fractions of droplets. Although it would be hard to quantify, it is likely that the soft layer of adsorbed proteins may undergo some changes at high volume fraction, such as deswelling or interpenetration.

\subsubsection{Protein suspensions}
\begin{figure}[htb]
	\begin{center}
		\includegraphics[width=0.8\textwidth]{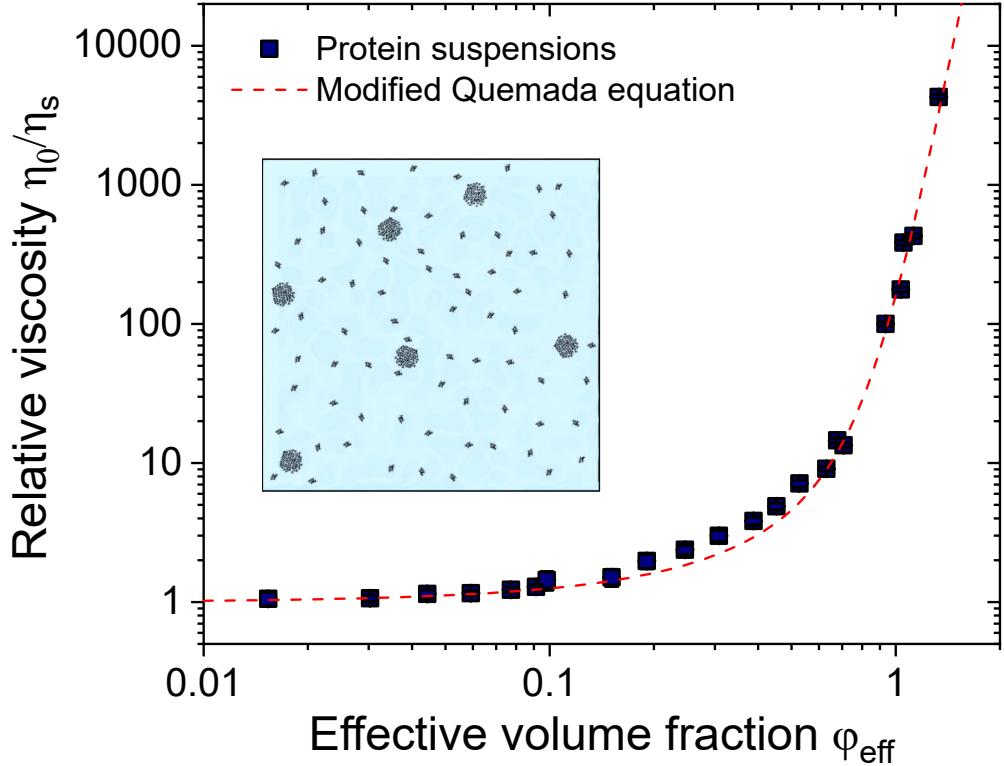}
	\end{center}
	\caption{Relative viscosity of sodium caseinate suspensions ($\square$, navy) as a function of the effective volume fraction. The red dashed line denotes the modified Quemada equation, Equation~\ref{Eq:ModifQuemadaSoft}, the values for $n$ and $\phi_m$ are listed in table~\ref{Tab:ModifQuemadaProtParameters}.}
	\label{Fig:ViscoProtSoftQuemada}
\end{figure}

Sodium caseinate is known to aggregate in solution to form clusters or micelles \cite{farrer:1999, pitowski:2008, huppertz:2017}. These differ from protein-stabilised droplets because of their swollen structure, and likely dynamic nature. The viscosity behaviour of the suspensions they form is displayed in  Figure~\ref{Fig:ViscoProtSoftQuemada}. 

At high concentrations, the viscosity does not diverge as quickly as for the suspensions of droplets. This result is in agreement with previous studies on sodium caseinate, in which suspensions at higher concentrations were studied \cite{farrer:1999,pitowski:2008,loveday:2010}. In these works, it was shown that the viscosity does not diverge but follows a power law $\eta_0/\eta_s\propto(\phi_{eff,prot})^{12}$ . 

The behaviour displayed by sodium caseinate resembles that of core-shell microgels \cite{tan:2005} and soft spherical brushes\cite{vlassopoulos:2001}, hence a soft colloid framework (as reviewed e.g. in Ref. \cite{vlassopoulos:2014}) seems suitable for the study of these suspensions. 

A general feature of the viscosity behaviour of soft colloidal suspensions is the oblique asymptote at high concentrations. This behaviour is beleived to arise because, as the concentration increases, the effective volume occupied by each particle decreases, by de-swelling or interpenetration. Thus, the strong viscosity divergence of hard-sphere suspensions is absent for soft colloids. To describe the behaviour of such suspensions, a model is thus required, that takes into account this distinctive limit at high concentrations while retaining the hard sphere behaviour at lower concentrations.

A semi-empirical modification that fulfills the above criteria is the substitution of the maximum volume fraction $\phi_m$ by a $\phi$-dependent parameter $\phi_m^*$ that takes the form: $\phi_m^* = \left({\phi_m}^n+\phi^n\right)^{1/n}$.

As a result, a modified version of Equation~\ref{Eq:QuemadaHS} can be derived, that takes into account the softness of the particles via a concentration-dependent maximum volume fraction $\phi_m^*$. This semi-empirical viscosity model is expressed:
\begin{equation}
\label{Eq:ModifQuemadaSoft}
\frac{\eta_0}{\eta_s} = \left(1-\frac{\phi}{\phi_m^*}\right)^{-2} 
\end{equation}
where:
\begin{equation*}
\phi_m^* = \phi_m\left(1+\left(\frac{\phi}{\phi_m}\right)^n\right)^{1/n}
\end{equation*}
The addition of the exponent $n$ as a parameter expresses the discrepancy from the hard-sphere model. The smaller $n$, the lower the volume fraction $\phi$ at which $\phi^*_m$ diverges from $\phi_m$, and the less sharp the divergence in viscosity.

The model in Equation~\ref{Eq:ModifQuemadaSoft} was applied to fit the experimental data displayed in Figure~\ref{Fig:ViscoProtSoftQuemada}, and the resulting fitting parameters are listed in Table~\ref{Tab:ModifQuemadaProtParameters}. 
\begin{table}[hbt]
\begin{ruledtabular}
	\centering
	\caption{Parameters for modified Quemada model for soft colloids, Equation~\ref{Eq:ModifQuemadaSoft}, applied to sodium caseinate suspensions}
	\begin{tabular}{ccc}

		Parameter & Value & Standard Error \\
		\hline
		$\phi_m$ & \num{0.93} & \num{0.02} \\
		$n$ & \num{6.1} & \num{0.4}\\

	\end{tabular}
	\label{Tab:ModifQuemadaProtParameters}
	\end{ruledtabular}
\end{table}
The use of this approach gives a good fit of the viscosity behaviour of sodium caseinate in the range of concentrations used here. In addition, this semi-empirical model also satisfactorily describes the viscosity of sodium caseinate suspensions at higher concentration from Ref.~\cite{farrer:1999,pitowski:2008}. It is worth noting that the inflection of viscosity is slightly sharper for the model than for the experimental data.

The power-law towards which the relative viscosity $\eta_0/\eta_s$ described by Equation~~\ref{Eq:ModifQuemadaSoft} tends at high concentration (ie $\phi>\phi_m$) can be calculated by developing $\phi_m^*$. Indeed, at high concentration $\phi_m^*$ converges towards $\phi\times\left(1+\frac{1}{n} \times \left(\frac{\phi_m}{\phi}\right)^n\right)\propto(\phi_{eff})^{n}$, so $\eta_0/\eta_s$ converges towards $\left(1+n\left(\frac{(\phi_{eff})}{\phi_m}\right)^{n}\right)^{2}$ (detailed calculations are provided in the supplementary material). 
Using the value in Table~\ref{Tab:ModifQuemadaProtParameters}, the relative viscosity of sodium caseinate suspensions is found to  follow the power law $\eta_0/\eta_s\propto(\phi_{eff,prot})^{\num[separate-uncertainty=true]{12.2(8)}}$. This value is in good agreement with the literature where $\eta_0/\eta_s\propto(\phi_{eff,prot})^{12}$ in the concentrated regime \cite{farrer:1999,pitowski:2008,loveday:2010}.

It is interesting to note that Equation~\ref{Eq:ModifQuemadaSoft} provides a good model for the behaviour of particle suspensions and emulsions whose particles have a wide range of softness, as will be detailed elsewhere. Within this context the concentration behaviour of sodium caseinate suspensions seems to indicate that they can also be regarded as suspensions of soft particles. This interpretation of the behaviour can be further tested by considering the shear-rate dependent response of both the emulsions and sodium caseinate suspensions. 

\subsection{Shear thinning behaviour of protein and droplet suspensions}
 Over most of the concentration range studied here, the protein suspensions display Newtonian behaviour. However, at high concentration of protein, shear thinning is observed at high shear-rates (flow curves in supplementary material). By comparison, the droplet suspensions display shear thinning at a much broader range of concentrations. This behaviour is common in colloidal suspensions \cite{dekruif:1985,helgeson:2007}, as well as polymer and surfactant solutions and arises from a variety of mechanisms \cite{cross:1965, cross:1970}. In non-aggregated suspensions of Brownian particles shear-thinning arises from the competition between Brownian motion (which increases the effective diameter of the particles) and the hydrodynamic forces arising from shear. Shear thinning then occurs over a range where the two types of forces balance, as characterised by the dimensionless reduced shear stress ($\sigma_{r}$) being of order unity. $\sigma_{r}$ is given by
\begin{equation}
\sigma_{r} = \frac{\sigma R^3}{k T}
\label{Eq:SigmaRC}
\end{equation}
where $R$ is the radius of the colloidal particle, $k$ is the Boltzmann constant and $T$ is the temperature of the suspension (here $T = \SI{298}{\kelvin}$). 

In such suspensions, the flow curve can be described using the following equation for the viscosity as a function of shear stress \cite{woods:1970, krieger:1972, frith:1987}:
\begin{equation}
\frac{\eta}{\eta_s} = \eta_\infty + \frac{\eta_0-\eta_\infty}{1+\left(\sigma_r/\sigma_{r,c}\right)^m}
\label{Eq:ShearThin}
\end{equation}
Where $\eta_0$ is the zero-shear viscosity, $\eta_\infty$ is the high-shear limit of the viscosity, $m$ is an exponent that describes the sharpness of the change in regime between $\eta_0$ and $\eta_\infty$, and $\sigma_{r,c}$ is the reduced critical shear stress. 

Because shear thinning arises from the competition between Brownian motion and the applied external flow, the use of a dimensionless stress that takes into account the size of the colloidal particles allows meaningful comparisons between the different suspensions\cite{woods:1970, frith:1987}. Here, we use this approach to compare the flow behaviour of the protein and droplet suspensions, and to test further the hypothesis that the protein suspensions can be considered to behave as though they are suspensions of soft particles.

\begin{figure}[htb]
	\includegraphics[width=0.9\textwidth]{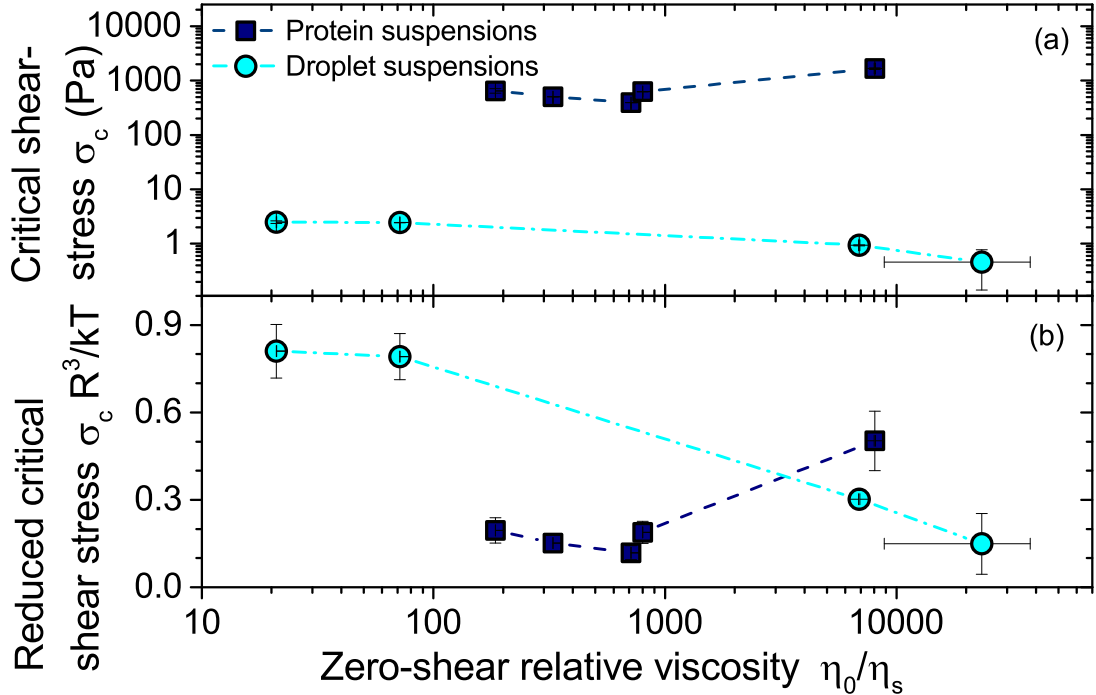}
	\caption{Shear thinning behaviour of concentrated suspensions of sodium caseinate ($\square$, navy), and sodium-caseinate stabilised droplets ($\circ$, cyan) as characterised by the critical shear stress for shear-thinning. (a) Critical shear stress $\sigma_c$ as a function of the zero-shear relative viscosity $\eta_0/\eta_s$ for several concentrated suspensions. $\sigma_c$  and $\eta_0$ were estimated by fitting the flow curves (Figure S1 in supplementary material) with Equation~\ref{Eq:ShearThin}. (b) Reduced critical shear stress $\sigma_{r,c}$ \ref{Eq:SigmaRC} as a function of the zero-shear relative viscosity $\eta_0/\eta_s$. The error bars indicate the uncertainty of the fitting parameters (more details are provided in the supplemetary material), and the lines are indicated as guide for the eye.}
	\label{Fig:SigmaCShearThin}
\end{figure}

Fitting the flow curves with Equation~\ref{Eq:ShearThin} allows for the extraction of the critical stress $\sigma_c$. The behaviour of this parameter as a function of the zero shear relative viscosity (as a proxy for concentration) is shown in Figure~\ref{Fig:SigmaCShearThin}(a). The corresponding values of $\sigma_{r,c}$  are calculated using $R_{drop} \equiv R_{h,drop} = \SI{110}{\nano\metre}$ and $R_{prot}\equiv R_{h,prot} = \SI{11}{\nano\metre}$, are displayed in Figure~\ref{Fig:SigmaCShearThin} (b). 

As can be observed, the protein suspensions require a much higher stress to produce a decrease in viscosity than do the droplet suspensions, as $\sigma_c$ is more than two orders of magnitudes higher. However, this difference is largely absent when the reduced critical shear-stress is used, indicating that the main difference between both systems is the size of the particles and that there are no differences in interparticle interactions at high concentrations, notably no further extensive aggregation of sodium caseinate. 

Shear thinning is thus another aspect of the rheology of sodium caseinate that shows an apparent colloidal behaviour rather than polymeric behaviour 
. This result reinforces the relevance of the soft colloidal framework as an approach for studying the viscosity of sodium caseinate and sodium caseinate-stabilised droplets. 

\section{Viscosity of mixtures}

After having studied separately the components of protein-stabilised emulsions, the next logical step is to investigate mixtures of both with well-characterised compositions by combining purified droplets and protein suspensions. In addition, the soft colloidal framework developed above provides a basis for the development of a predictive approach to the viscosity of mixtures of proteins and droplets, as formed upon emulsification of oil in a sodium caseinate suspension. These topics are the subject of the current section.

\begin{figure}[htb]
	\begin{center}
		\includegraphics[width=0.7\textwidth]{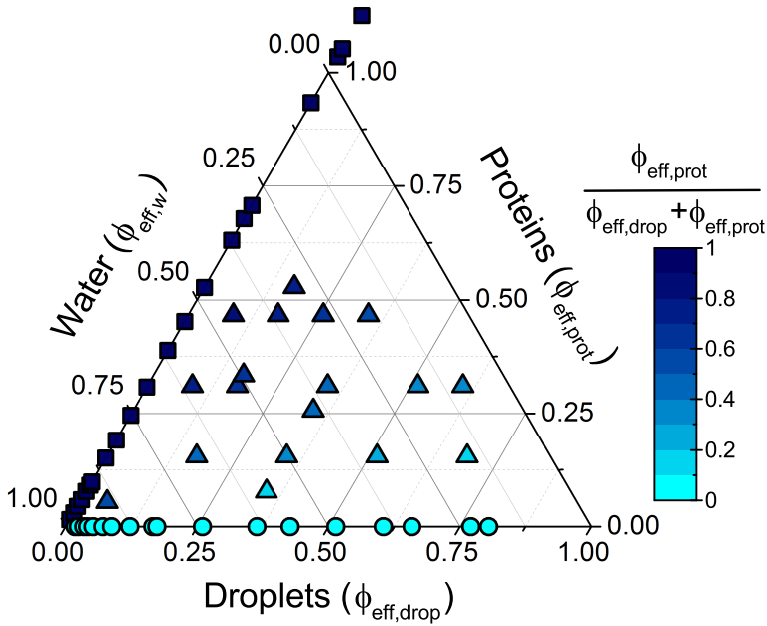}
	\end{center}
	\caption{Composition of suspensions of sodium caseinate ($\square$, navy), sodium-caseinate stabilised droplets ($\circ$, cyan), and of mixtures ($\triangle$, colour-coded as a function of $\chi_{prot}$ defined in Equation~\ref{Eq:ChiProt}). 
	}
	\label{Fig:SamplesVisco}
\end{figure}

These mixtures are composed of water and of two types of colloidal particles (droplets and protein aggregates), hence they are conveniently represented as a ternary mixture, as displayed on Figure~\ref{Fig:SamplesVisco}. This representation is limited by the high volume fractions reached by proteins in suspension, hence  some data points lie outside of the diagram. The two-dimensional space of composition for the mixtures can be described by the total effective volume fraction $\phi_{eff,tot} = \phi_{eff,prot}+\phi_{eff,drop}$ and the ratio of their different components $\chi_{prot}$:
\begin{equation}
\label{Eq:ChiProt}
\chi_{prot}=\frac{\phi_{eff,prot}}{\phi_{eff,prot}+\phi_{eff,drop}}
\end{equation}
$\chi_{prot}$ describes the relative percentage of protein in the emulsion compared to the droplets: $\chi_{prot}=1$ for samples containing only proteins, $\chi_{prot}=0$ for samples containing only protein-stabilised droplets, and $\chi_{prot}=0.5$ for mixtures containing an equal volume fraction of proteins and protein-stabilised droplets.

\begin{figure}[htb]
	\begin{center}
		\includegraphics[width=0.9\textwidth]{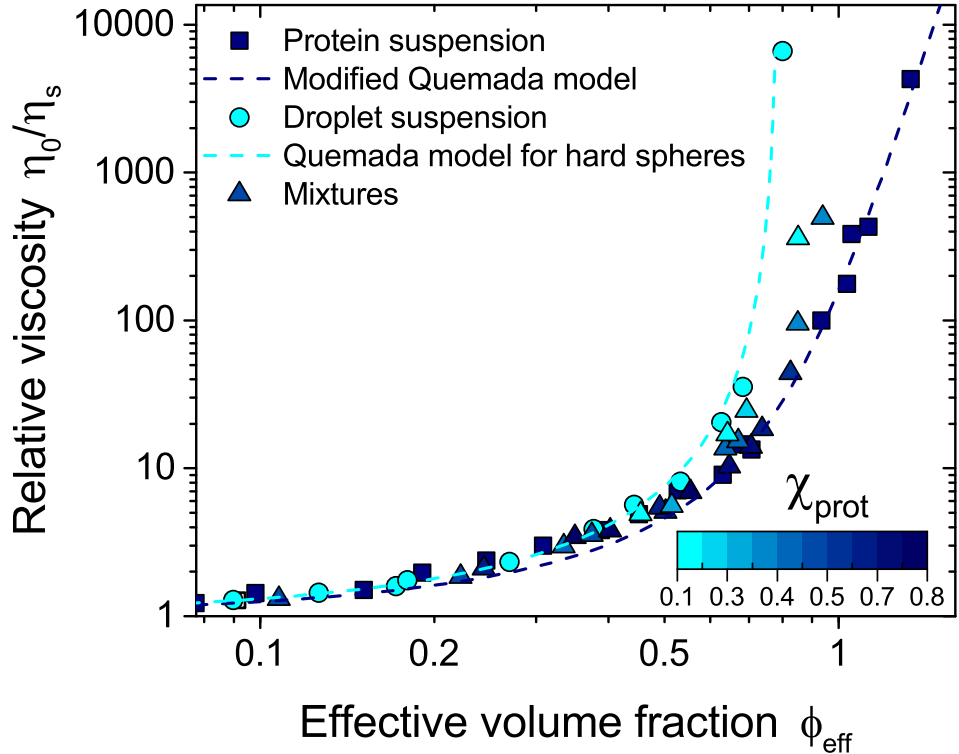}
	\end{center}
	\caption{Relative viscosities $\eta_0/\eta_s$ of suspensions as a function of the effective volume fraction $\phi_{eff}$: sodium caseinate suspensions ($\square$, navy), sodium-caseinate stabilised droplets suspensions ($\circ$, cyan), and  suspensions of mixtures ($\triangle$, colour-coded as a function of $\chi_{prot}$ defined in Equation~\ref{Eq:ChiProt}).}
	\label{Fig:RawViscoMix}
\end{figure}

The viscosity of the mixtures containing both proteins and protein-stabilised droplets was measured as for the pure suspensions. The values can be compared with the pure suspensions using the total volume fraction for the mixtures $\phi_{eff,tot}$, and are displayed in Figure~\ref{Fig:RawViscoMix}.

The mixtures all display viscosities between those of the pure droplets and of the pure proteins at a given volume fraction, their exact value depending on their compositional index $\chi_{prot}$.  
Notably, no phase separation is observed in the emulsion samples on the timescale of the experiments. This is an unusual result as sodium caseinate-stabilised emulsions are notoriously prone to depletion induced-flocculation caused by the presence of unadsorbed sodium caseinate \cite{bressy:2003,srinivasan:1996,dickinson:1997,dickinson:2010,dickinson:1999}. Presumably, this unusual behaviour is due to the small size of the droplets, which are only one order of magnitude larger than the naturally-occurring caseinate structures.

The knowledge and models introduced for the suspensions of proteins and droplets in the previous sections can be used to develop a semi-empirical model to describe the viscosity of mixtures.

\subsection{Semi-empirical predictive model}

Models have been developed previously to predict the viscosity of suspensions of multi-modal particles, for example in references~\cite{mendoza:2017} or~\cite{mwasame:2016a}, the latter was then extended for mixtures of components of different viscosity behaviours in~\cite{mwasame:2016b}. However these models are mathematically complex and do not describe accurately our experimental results.

\begin{figure}[htb]
	\centering
	\includegraphics[width=0.9\textwidth]{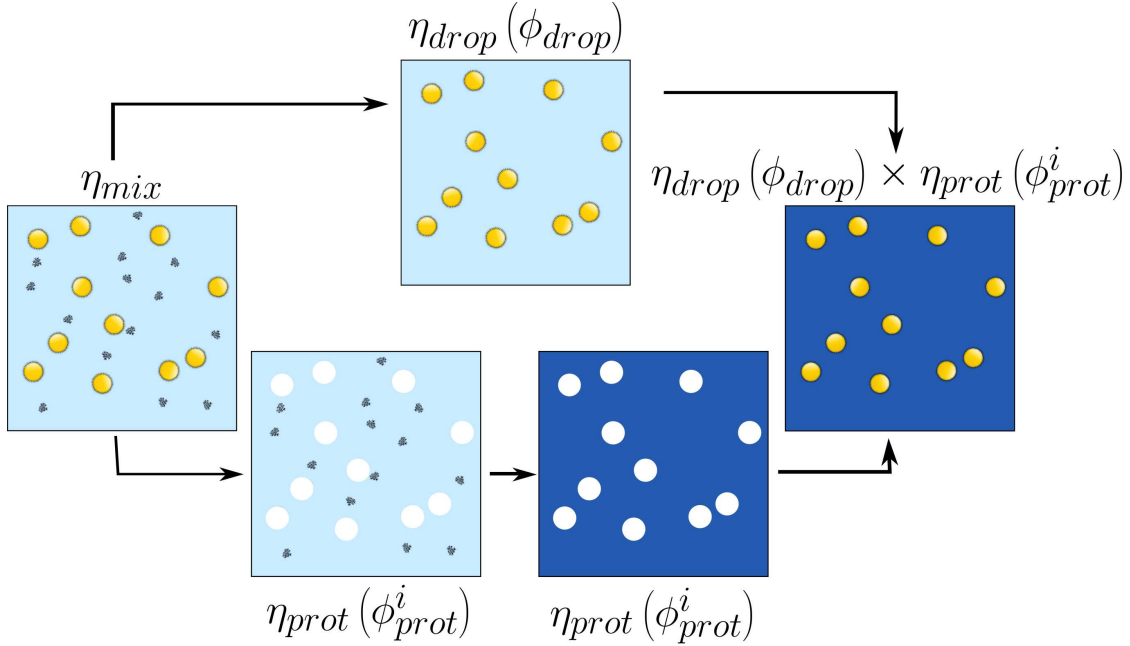}
	\caption{Development of a semi-empirical model to predict the viscosity of emulsions. The contribution of the proteins in suspension to the viscosity of the emulsion is modelled by an increase of viscosity of the continuous medium.}
	\label{Fig:MixViscoSchema}
\end{figure}

Instead, a simple and useful approach is to consider that each component of the mixture is independent from the other, as in the early model for multi-modal suspensions described in~\cite{farris:1968}. In this case, the protein suspension acts as a viscous suspending medium for the droplets, whose viscosity behaviour was previously characterised and modelled by Equation~\ref{Eq:QuemadaHS}. Because the viscosity behaviour of the protein suspension is also known, it can be combined with the droplet behaviour to determine the viscosity of the mixture. This approach is illustrated on Figure~\ref{Fig:MixViscoSchema}.

\subsubsection{Development of the model}

Considering the suspending medium alone first, it is useful to consider the protein content of the aqueous phase residing in the interstices between the droplets, $\phi_{prot}^i$:
\begin{equation}
\label{Eq:PhiProtInter}
\phi_{prot}^i=\frac{V_{prot}}{V_{prot}+V_{water}}=\frac{\phi_{prot}}{\phi_{prot}+\phi_{water}}=\frac{\phi_{prot}}{1-\phi_{droplet}}
\end{equation}
Where it is assumed that $\phi_{prot}\simeq \phi_{eff,prot} = k_{0,prot}\times c_{prot}$ and $\phi_{droplet} \simeq \phi_{eff,drop} = k_{0,drop}\times c_{drop}$ according to Eq.~\ref{Eq:EffPhi_proportional}, with $k_{0,prot}$ and $k_{0,drop}$ determined previously using the Batchelor equation fitted to the viscosities of semi-dilute suspensions of pure proteins and pure droplets. 

The study of the pure suspensions of protein-stabilised droplets and of proteins makes it possible to model the viscosity behaviour of both suspensions:
\begin{itemize}
	\item The relative viscosity of a suspension of protein-stabilised droplets $\eta_{r,drop}(\phi)$ is described by Equation~\ref{Eq:QuemadaHS} with the parameter $\phi_{m}=\num[separate-uncertainty=true]{0.79(2)}$ (Quemada model for hard spheres \cite{quemada:1977})
	\item The relative viscosity of a suspension of sodium caseinate $\eta_{r,prot}(\phi)$ is described by Equation~\ref{Eq:ModifQuemadaSoft} with the parameters listed in Table~\ref{Tab:ModifQuemadaProtParameters} (modified Quemada model) and using $\phi_{prot}^i$ as described above.
\end{itemize} 

These elements are then combined to predict the relative viscosity of the mixture $\eta_{r,mix}^p$, in the absence of specific interactions between the droplets and the proteins, thus: 
\begin{equation}
\label{Eq:ViscoMixModel}
\eta_{r,mix}^p(\phi_{eff,prot}, \phi_{eff,drop}) = \eta_{r,prot}\left(\phi_{prot}^i\right) \times \eta_{r,drop}\left(\phi_{eff,drop}\right)
\end{equation}

\subsubsection{Application of the model}
The values of the relative viscosity calculated for each mixture using Equation~\ref{Eq:ViscoMixModel} are compared to the experimentally measured relative viscosity $\eta_{r,mix}^m$, in Figure~\ref{Fig:RawViscoMix}. Details of the estimated viscosity of the continuous phase of the mixture can be found in the supplementary material (Figure S3).

\begin{figure}[htb]
	\includegraphics[width=0.8\textwidth]{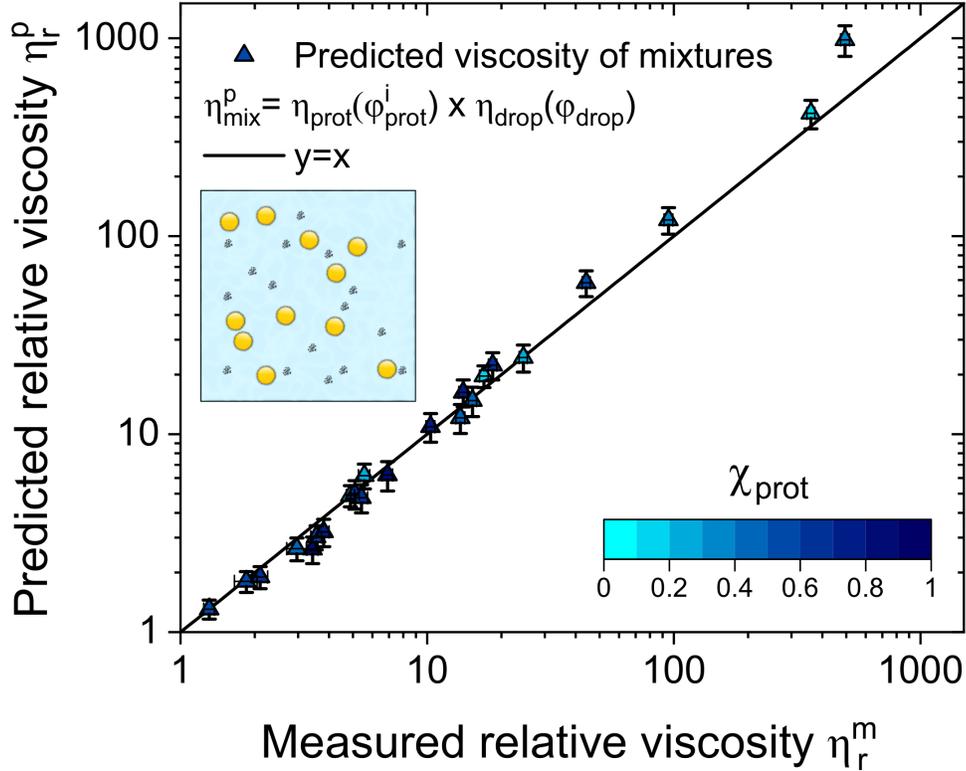}
	\caption{Predicted relative viscosity of mixture suspensions $\eta_{r,mix}^p$, calculated with Equation~\ref{Eq:ViscoMixModel}, as a function of the measured viscosity $\eta_{r,mix}^m$ from Figure~\ref{Fig:RawViscoMix}. Each point is a mixture of different composition
		, and its colour indicates the value of the compositional index $\chi_{prot}$ defined by Equation~\ref{Eq:ChiProt}. The straight line represents y=x. The error bars indicate the uncertainty arising from the calculations (more details are provided in the supplemetary material).}
\end{figure}

Despite the simplicity of this model, it provides a reasonably accurate prediction of the viscosity of protein-stabilised emulsions. This result seems to indicate that there are no specific interactions between the proteins and the droplets, neither at a molecular scale between un-adsorbed and adsorbed proteins, nor at a larger length scale where depletion interactions could occur. This is likely to be related to the small size of the droplets in this specific system, and increasing the droplet size may result in a decreased accuracy of this simple model. 

The small inaccuracies in the predicted viscosities probably lie in the slightly imperfect fit of Equations ~\ref{Eq:QuemadaHS} and ~\ref{Eq:ModifQuemadaSoft}. First, at moderate viscosity ($\eta_r<10$), the slight discrepancy between predicted and measured viscosity of the samples with a high $\chi_{prot}$ is probably a reflection of the modest underestimation of the viscosity of protein suspensions for $2<\phi_{eff}<10$ by Equation~\ref{Eq:ModifQuemadaSoft}. 

At higher concentrations, the effective volume fraction approximation may break down. Indeed, as observed previously for pure suspensions, $\phi_{eff}$ can reach high values and may not correspond exactly to the volume fraction actually occupied by the particles, especially in the case of $\phi_{eff,prot}$ . A natural consequence is that the relationship $\phi_{eff,prot} + \phi_{eff,drop} + \phi_{eff,water} = 1$ may not be verified, leading to an overestimation of $\phi_{prot}^i$ when calculated by Equation~\ref{Eq:PhiProtInter}. It should be noted that the lack of unifying definition of the volume fraction for soft colloids is a particularly relevant challenge when dealing with mixtures. An approach to address this problem could be to take the viscosity behaviour of one of the two components as a reference, and map the volume fraction of the other component to follow this reference viscosity \cite{mwasame:2016b}, but it would considerably increase the complexity of the model.

Finally, another possible source of discrepancy is the assumption that the proteins in the interstices will reach the same random close packing fraction as for proteins in bulk $\phi_{rcp,prot}$. However, at high droplet volume fraction, there are geometrical arguments to support the hypothesis of a different random close packing volume fraction due to excluded volume effects. Therefore, this assumption may lead to a decreased accuracy of the model at high concentrations.

To summarize, in this section we have shown that the preliminary study of the individual components of a mixture allows the subsequent prediction of the viscosity of mixtures of these components with reasonable accuracy, providing that the composition of the mixtures is known.

\subsubsection{Reversal of the model: estimation of the composition of emulsions}

\begin{figure}[htb]
	\includegraphics[width=0.9\textwidth]{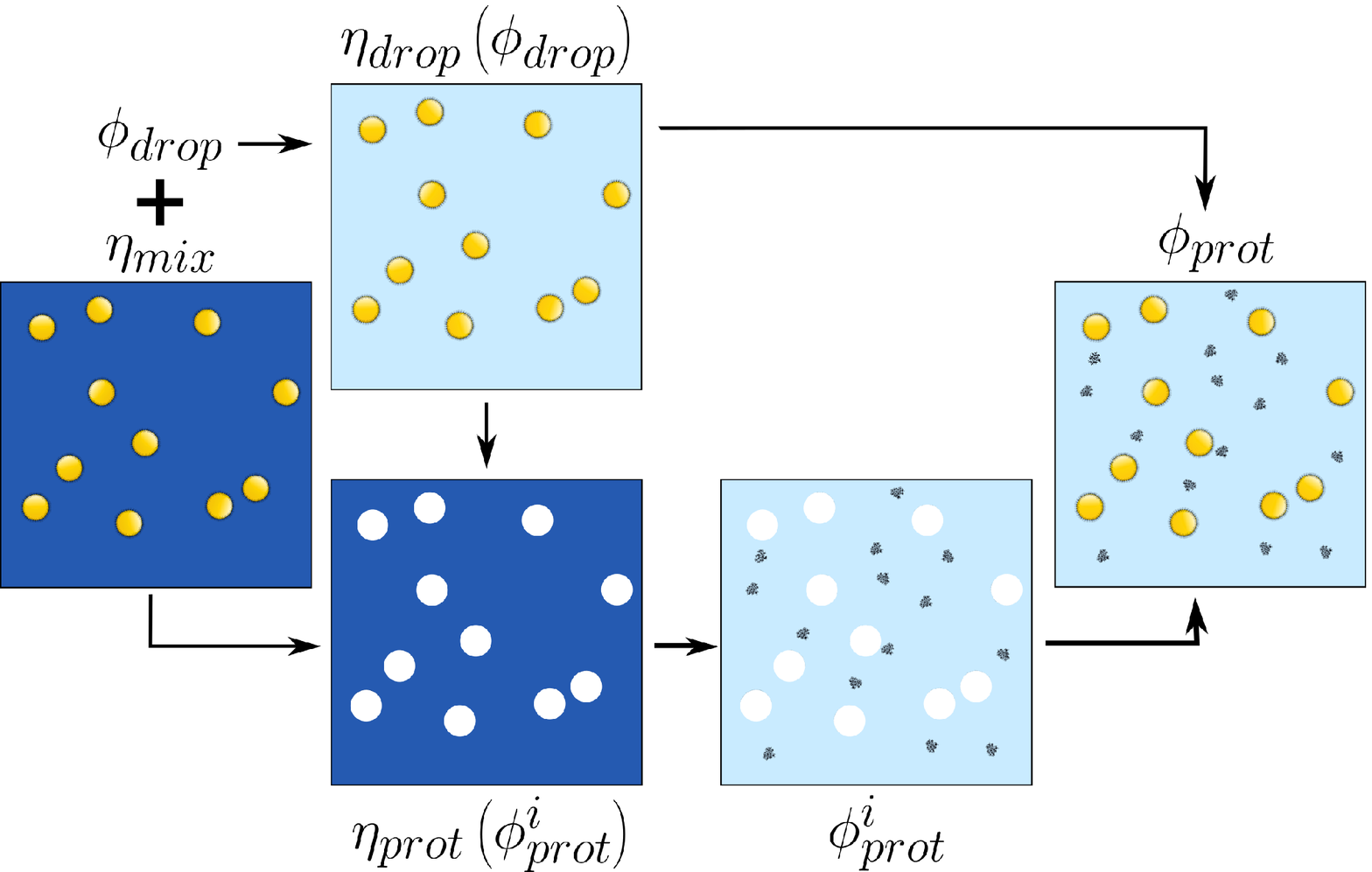}
	\caption{Reversal of semi-predictive model for the viscosity of protein-stabilised emulsions. The measurement of the emulsion viscosity $\eta_{r,mix}$ makes possible the calculation of the volume fraction of un-adsorbed proteins $\phi_{eff,prot}$, given that the volume fraction of droplets $\phi_{eff,drop}$ is known from the preparation protocol.}
	\label{Fig:PredictViscoReverse}
\end{figure}

A common challenge when formulating protein-stabilised emulsions is to estimate the amount of protein adsorbed at the interface as opposed to the protein suspended in the aqueous phase. Here we suggest that reversing the semi-empirical model developed in the previous section allows estimation of the amount of proteins in suspension after emulsification with a simple viscosity measurement, which can be performed on-line in advanced industrial processing lines. The calculation process is illustrated in Figure~\ref{Fig:PredictViscoReverse}.

To assess the accuracy of the suggested method, a case in point is the emulsion used to prepare the sodium caseinate droplets in this study after microfluidisation. It is composed of $\num{20}\%$(wt) oil and $\num{4.0}\%$(wt) sodium caseinate, and its relative viscosity was measured to be $\eta_{r,mix}^m=\num{10}$.

The first step is to calculate the contribution of the oil droplets to the viscosity of the mixture, in order to isolate the protein contribution. A $\num{20}$(wt)$\%$ content in oil corresponds to $\phi_{eff,drop}=\num{0.40}$, so $\eta_{r,drop}=\left(1-\phi_{eff,drop}/\phi_m\right)^{-2}=\num{4.1}$.

It is then possible, using the Equation~\ref{Eq:ViscoMixModel}, to calculate the viscosity of the continuous phase $\eta_{r,prot}\left(\phi_{prot}^i\right)=\eta_{r,mix}^m/\eta_{r,drop}=2.4$, assumed to arise from the presence of un-adsorbed proteins. In order to estimate the volume fraction of proteins in the interstices $\phi_{prot}^i$, the equation below has to be solved:
\begin{equation}
\label{Eq:MixReverse}
\left(1+\left(\frac{\phi_{eff,prot,m}}{\phi_{prot}^i}\right)^n\right)^{-1/n}=1-\frac{1}{\sqrt{\eta_{r,prot}}}
\end{equation}

Finally, numerically solving Equation~\ref{Eq:MixReverse} with the values for $n$ and $\phi_m$ from Table~\ref{Tab:ModifQuemadaProtParameters} gives $\phi_{prot}^i=\num{0.33}$. This result corresponds to a volume fraction of un-adsorbed proteins in the overall emulsion $\phi_{eff,prot} = \phi_{prot}^i (1-\phi_{eff,drop})=\num{0.20}$, or expressed as a concentration in the emulsion: $c=\SI{23}{\milli\gram\per\milli\liter}$. This has to be compared with the initial concentration of $\SI{45}{\milli\gram\per\milli\liter}$ in proteins before emulsification. Thus, only half of the amount of proteins adsorb at the interface, while the other half is still in suspension.

This result can be converted into a surface coverage to be compared with studies on sodium caseinate-stabilised emulsions using micron-sized droplets. It is estimated that $\SI{1}{\liter}$ of emulsion containing  $\num{20}$(wt)$\%$ of oil, and with a droplet size of $R_{opt,c}=\SI{65}{\nano\metre}$ presents a surface area of oil of $\SI{920}{\metre\squared}$, and from the viscosity $\SI{22}{\gram}$ of sodium caseinate is adsorbed at the interfaces. Thus, the surface coverage is around $\SI{24}{\milli\gram\per\meter\squared}$. This result is in good correspondence with studies on similar emulsions at larger droplet sizes \cite{srinivasan:1996, srinivasan:1999}, and thus provides a validation for the use of the measurement of the viscosity as a tool to estimate the amount of unadsorbed proteins present in emulsions.

The semi-empirical model for the viscosity of emulsions developed in this study, once calibrated, can thus be used not only as a predictive tool for mixtures of droplets and proteins of known composition, but also as a method to estimate the amount of adsorbed proteins without the need for further separation of the components.

\section{Conclusion}

Previous studies have attempted to compare the rheological properties of sodium caseinate to those of a suspension of hard spheres, and found that agreement at high concentrations is poor\cite{farrer:1999, pitowski:2008}. As a result it was concluded that a colloidal model is inadequate to describe the observed behaviour. Here we argue that this is mainly due to the choice of hard spheres as colloidal reference. We have shown that using the framework developed for soft colloidal particles, such as microgels and block co-polymer micelles \cite{vlassopoulos:2014}, helps toward a better description of the viscosity behaviour of the protein dispersions. Although this approach neglects the additional layer of complexity due to the biological nature of the sodium caseinate, such as inhomogeneous charge distribution and dynamic aggregation \cite{sarangapani:2013, sarangapani:2015}, it gives a satisfactory model that can be used to build a better description of protein-stabilised emulsions. Interestingly, the soft colloidal approach can also be successfully applied to the rheology of non-colloidal food particles, such as fruit purees \cite{leverrier:2017}.

In addition, a protocol was developed for preparing pure suspensions of protein-stabilised droplets rather than emulsions containing unadsorbed proteins. The viscosity behaviour of the nano-sized droplets appeared to be very similar to the hard sphere model. The main discrepancy is the high effective volume fraction at which the viscosity diverges, which may be due to the size distribution of droplets or arise from the softness of the layer of adsorbed proteins.

Finally, examining protein-stabilised emulsions as ternary mixtures of water, unadsorbed proteins and droplets has allowed us to develop a semi-empirical model for their viscosity. The contributions of each component to the overall viscosity of the emulsions being quantified by the analysis of the properties of the pure suspensions of droplets or proteins. The model can also be reversed to estimate the composition, after emulsification, of a protein-stabilised emulsion given its viscosity. It should be noted, however, that the droplet size is likely to be critical to the success of the model, as it is expected that flocculation of droplets will occur for larger droplets \cite{bressy:2003,srinivasan:1996,dickinson:1997,dickinson:2010,dickinson:1999}. This is due to the depletion interaction generated by the proteins in the mixture, which is not taken into account in the present model. For this reason, it would be interesting to explore further the influence of the droplet size on the viscosity behaviour of emulsions. In addition, increasing the droplet size would change the hardness of the droplets by decreasing the internal pressure as well as the influence of the soft layer of proteins, adding further complexity to the system.

\section*{Supplementary material}
The Supplementary material contains information on the calculation of the error bars, the viscosity as a function of the concentration, calculations of the asymptotic behaviour of Equation~\ref{Eq:ModifQuemadaSoft}, flow curves of the shear-thinning samples and the contributions to the viscosity of mixtures by the dispersed and continuous phases.

\begin{acknowledgments}
This project forms part of the Marie Curie European Training Network COLLDENSE that has received funding from the European Union’s Horizon 2020 research and innovation programme Marie Sk\l{}odowska-Curie Actions under the grant agreement No. 642774. The authors wish to acknowledge DMV for graciously providing the sodium caseinate sample used in this study, and PostNova Analytics Ltd for graciously performing the field flow fractionation measurement of sodium caseinate.
\end{acknowledgments}

\bibliography{References}

\end{document}